# A Taxonomy of Snow Crystal Growth Behaviors: 2. Quantifying the Nakaya Diagram

Kenneth G. Libbrecht

Department of Physics, California Institute of Technology
Pasadena, California 91125, kgl@caltech.edu

**Abstract.** This paper presents a matrix of 206 snow crystal growth observations as a function of temperature and water vapor supersaturation in air, each illustrating the morphology and size of a crystal forming on the tip of an isolated c-axis ice needle after a known growth time. Because each complex structure emerged from a simple, well-defined seed crystal under known environmental conditions, this data set is well suited for making comparisons with three-dimensional computational models. These observations thus provide a needed extension of the well-known Nakaya diagram, as they allow a quantitative evaluation of model predictions over a broad range of growth conditions. I also briefly discuss computational methods along with an initial model of the most relevant microphysical processes governing snow crystal growth. My overarching goal with this new data set is to facilitate the development of quantitative computational growth models that can eventually reproduce the remarkable diversity of morphological structures seen in snow crystal formation.

## ❅ Developing realistic computational models of snow crystal growth

Being an intrinsically complex phenomenon, snow crystal formation cannot be described by a basic analytic theory, at least not in the simplest sense of the word. Instead, it is necessary to first break the problem down into its constituent parts, characterizing the relevant physical processes acting over different length- and time- scales. Then one reassembles those parts into a computational model to replicate the full range of complex growth behaviors. If the model is physically sensible and agrees with quantitative experimental observations, then we can rightly say that we generally understand the phenomenon [2021Lib, 2017Lib]. At present, however, we are far from achieving this goal, and the remarkable morphological diversity observed in snow crystal growth remains generally inexplicable.

Creating a suitable computational model of snow crystal growth presents a challenging technical task, partly because the underlying molecular dynamics of ice solidification is itself complex and quite subtle, and partly because modeling highly anisotropic attachment kinetics during crystal growth can lead to undesirable computational instabilities. As a result, physically realistic models of structure formation that simultaneously exhibit both faceting and branching behaviors do not yet exist, and the snow-crystal problem has proven especially troublesome.

The overarching topic of structure formation during solidification has received much attention in the scientific literature, and numerous reviews are available [2017Jaa, 2002Boe, 2016Kar, 2018Che]. Much of this work focuses on systems with generally weak anisotropies in both the attachment kinetics and surface energies, however, as these systems are more amenable to computational modeling than systems exhibiting highly anisotropic faceting like snow crystals.

Focusing on the snow crystal problem, Barrett, Garcke, and Nürnberg recently presented numerical simulations of growing snow crystals using a finite-element method in which the ice surface was approximated using an adaptive polygonal mesh [2012Bar]. This technique demonstrates one variant of a front-tracking strategy, as it defines a sharp solidification front between solid ice and the water-vapor field surrounding it [1996Sch, 2010Bar]. While reproducing some features seen in laboratory snow crystal growth, this model includes a highly anisotropic ice/vapor surface energy, which is physically unrealistic [2012Lib2]. Whether a similar front-tracking strategy can be developed using highly anisotropic attachment kinetics remains an open question.

More recently, Demange, Zapolsky, Patte and Brunel demonstrated a phase-field technique for simulating snow crystal growth [2017Dem, 2017Dem1], and this also reproduced several growth behaviors known from laboratory observations. Once again, however, the model used a highly anisotropic surface energy to produce faceted structures, which is physically unrealistic. And again, it is not clear if this model strategy can be applied to the case of highly anisotropic attachment kinetics, where numerical instabilities can be more troublesome.

Perhaps the most promising computational technique yet demonstrated for snow-crystal growth is that using *cellular automata* (CA) on a fixed crystalline grid, as this method can handle highly anisotropic attachment kinetics with relative ease. After initial demonstrations by Reiter and Ning [2005Rei, 2007Nin], Gravner and Griffeath greatly expanded these ideas in a series of influential papers [2006Gra, 2008Gra, 2009Gra], the latest demonstrating a full 3D snow crystal simulator exhibiting a remarkable diversity of realistic morphologies, including details that had hitherto not been seen in any numerical simulations.

While the Gravner and Griffeath papers used somewhat ad hoc CA propagation rules to describe the surface boundary conditions, Libbrecht showed that physically derived rules could be obtain directly from models of various surface processes [2008Lib, 2013Lib1, 2015Lib1], thus incorporating surface attachment kinetics, surface energy effects, and even surface diffusion effects into the CA formalism. Libbrecht et al. further demonstrated how 2D cylindrically symmetrical CA models could be compared with quantitative experimental data to good effect [2015Lib2].

Kelly and Boyer further developed these ideas into a fully physically derived 3D CA model for snow crystal growth [2014Kel], and this model may be the closest representation to date of actual snow-crystal formation. [Brief technical aside: In the CA model, it is not necessary to "invert" the function $\alpha(\sigma_{surf})$ as was suggested in [2014Kel]. Rather one can numerically relax the external supersaturation field while simultaneously relaxing the boundary values of $\sigma_{surf}$ and $\alpha(\sigma_{surf})$ at each point on the ice surface [2008Lib, 2013Lib1, 2015Lib1]. In this way, only the direct function $\alpha(\sigma_{surf})$ is needed].

Given the substantial amount of recent progress on this problem, it may indeed soon be possible to create physically realistic computational models of snow crystal formation over a broad range of growth conditions, thereby passing an important milestone in our understanding of this enigmatic phenomenon. To authenticate such a claim, however, it is necessary to go beyond demonstrating model morphological structures that qualitatively resemble snowflake photographs. Developing a physically realistic model of natural snow crystal growth requires a quantitative reproduction of observed morphologies and growth rates as a function of all relevant extrinsic variables, including temperature, supersaturation, background air pressure, etc. This is a high bar to clear, however, requiring:

1) A stable 3D computational method including a suitable parameterization of all relevant physical processes guiding crystal growth.
2) A reasonably accurate microphysical model and general understanding of these physical processes over the relevant length- and time- scales.
3) A comprehensive set of quantitative experimental data describing snow crystal growth rates and morphological behaviors over a broad range of growth conditions, which can be used to validate the computational models.

Other than the brief introduction above, the present paper will not focus on item (1) in this list, as computational modeling of crystal growth dynamics exhibiting both faceting and branching behaviors is itself a rich and fascinating subject. Rather, I will first focus on item (2) in the following section, outlining one approach to developing a suitable physical model of snow crystal growth dynamics, including recent work describing structure-dependent attachment kinetics (SDAK) and its effects on snow crystal growth [2019Lib1, 2021Lib]. Following that discussion, I move on to item (3) by presenting a set of experimental observations of snow crystal growth on the ends of c-axis ice needles, as this systems seems well suited for quantitative comparison with computational models.

# ❄ Developing a physical model of snow crystal growth dynamics

As described in [2021Lib, 2017Lib], the physical processes that are most important in snow crystal growth can be isolated and studied individually to some degree, following the time-honored tenets of scientific reductionism. Researchers in these areas are not always in uniform agreement on all topics, however, so I will next summarize the different physical effects with a slant toward my specific beliefs, appreciating that these views may require modification with future experimental and theoretical developments. Listing the relevant physical processes yields, in rough order of importance:

## 1) Particle diffusion

Particle diffusion is clearly an important factor governing snow crystal growth in normal air. The diffusion equation can be applied in the Laplace approximation because the supersaturation field equilibrates on timescales much shorter than typical crystal growth times [2021Lib]. The good news here is that the underlying physics of particle diffusion is well understood, and all computational techniques have no trouble solving the diffusion equation describing water vapor in air to high accuracy. The troubles arise mainly with determining the boundary conditions at the crystal surface.

## 2) Attachment kinetics

This moniker is given to the slate of physical processes that determine how water vapor molecules attach to ice surfaces under different conditions. It is customary to use the Hertz-Knudsen relation to write [2021Lib, 2017Lib]

$$v_n = \alpha v_{kin} \sigma_{surf} \qquad (1)$$

where $v_n$ is the crystal growth velocity perpendicular to the growing surface, $\alpha$ is a dimensionless *attachment coefficient*, $\sigma_{surf} = (c_{surf} - c_{sat})/c_{sat}$ is the water vapor supersaturation at the surface, $c_{surf}$ is the water-vapor number density immediately above the surface, $c_{sat} = c_{sat}(T)$ is the saturated number density of a surface in equilibrium at temperature $T$, and

$$v_{kin} = \frac{c_{sat}}{c_{ice}} \sqrt{\frac{kT}{2\pi m_{mol}}} \qquad (2)$$

is the *kinetic velocity*, in which $m_{mol}$ is the mass of a water molecule, $c_{ice} = \rho_{ice}/m_{mol}$ is the number density of ice, and $\rho_{ice}$ is the mass density of ice.

The relevant surface physics involved in snow crystal growth is incorporated into the attachment coefficient $\alpha$, whose value lies between zero and one. One can think of $\alpha$ as a sticking probability, equal to the fraction of water vapor molecule striking the ice surface that become immediately assimilated into the crystal lattice. The value of $\alpha$ may depend on $\sigma_{surf}$, $T$, surface orientation relative to the crystal axes, background gas pressure, surface chemical effects, and perhaps other factors.

Molecularly "rough" ice surfaces typically exhibit $\alpha_{rough} \approx 1$, as water vapor molecules striking a rough surface are usually immediately indistinguishable from those in the existing ice lattice. Meanwhile, it is common to find $\alpha_{facet} \ll 1$ on "smooth" faceted surfaces, as these have fewer open molecular binding sites, reducing the average sticking probability.

The overall basal/prism aspect ratio of a growing snow crystal is largely determined by the anisotropy of the attachment kinetics [2021Lib]. For example, thin snow-crystal plates only form when $\alpha_{basal} \ll \alpha_{pri}$, while slender columns only appear when $\alpha_{prism} \ll \alpha_{basal}$. Neither diffusion-limited growth nor surface-energy effects produce growing snow crystals with high overall aspect ratios.

Note that Equation 1 represents a purely local model of the attachment kinetics, in that the growth rate $v_n$ at each point on the surface derives solely from the values of $\sigma_{surf}$, $T$, and $\alpha$ at that point. A local model seems to work fairly well in the snow-crystal case because the molecular physics determining $\alpha$ occurs on length scales far smaller than other scales in the problem. Thus Equation 1 allows us to separate the long-range effects of water-vapor particle diffusion from the short-range effects of attachment kinetics. For the case of a flat surface of infinite extent, all non-local processes can be incorporated into $\alpha$, as then Equation 1 does little more than define $\alpha(\sigma_{surf}, T)$ from $v_n(\sigma_{surf}, T)$ in that ideal case.

That being said, a purely local model of the attachment coefficient may not adequately describe the attachment kinetics in all circumstances. For example, surface diffusion is an intrinsically non-local process that can alter the effective attachment coefficient, depending on surface morphology and other factors. Interestingly, non-local processes like surface diffusion play an especially large role in computational models of snow crystal growth, as the spatial resolution of the numerical grid is typically orders of magnitude larger than the molecular scale [2021Lib]. Given this mismatch in resolution, it remains to be seen if the concept of a purely local attachment coefficient is adequate for accurately modeling the molecular physics underlying the attachment kinetics for snow crystal growth.

### 3) Terrace nucleation

For the growth of faceted basal and prism surfaces, the attachment kinetics is well described by a terrace nucleation model [2021Lib]. For both facets, we write the attachment kinetics in the form

$$\alpha(T, \sigma_{surf}) = A e^{-\sigma_0/\sigma_{surf}} \quad (3)$$

where $\sigma_0(T)$ and $A(T, \sigma_{surf})$ are dimensionless parameters determined by laboratory ice-growth measurements [2021Lib, 2017Lib, 2013Lib]. Over the temperature range $0\,C > T > -30\,C$, the dependence of $\alpha(T, \sigma_{surf})$ on $\sigma_{surf}$ is dominated by the exponential factor, so we usually ignore any weak dependence of $A$ on supersaturation. Additionally, $\sigma_0(T)$ and $A(T)$ are generally different for the basal and prism facet surfaces.

### 4) Structure-dependent Attachment kinetics (SDAK)

The SDAK hypothesis posits that the attachment kinetics on broad faceted ice surfaces is intrinsically different from that on narrow facets [2021Lib]. We developed this hypothesis to explain the marked differences in ice growth behavior at low- and high-background gas pressures [2003Lib1] and advanced a molecular model of the SDAK phenomenon in [2021Lib, 2019Lib1].

Although the underlying physical assumptions in this model remain somewhat speculative, the SDAK phenomenon provides the only viable option currently available that can adequately explain the Nakaya diagram together with a plethora of other ice-growth data. The underlying SDAK molecular physics is at least plausibly realistic, and the model has been supported by several targeted experimental investigations [2020Lib1, 2020Lib2]. Thus, I am growing more confident that the essential elements of this model will withstand the test of time.

We still treat the SDAK phenomenon as a working hypothesis, and one of the best ways to further test this hypothesis is by comparing computational growth models with laboratory observations. Facilitating this avenue of research is the principal motivation behind the work presented here.

## 5) Surface energy effects

Although the surface-energy anisotropy in ice has not been definitively measured, the evidence suggests that it is small, and that the equilibrium shape of an isolated ice crystal is nearly spherical [2012Lib2]. The surface-energy anisotropy is thus negligible when calculating snow crystal growth dynamics, as it is completely dwarfed by the extremely large anisotropy in the attachment kinetics.

The (essentially isotropic) surface energy of the ice/vapor interface mainly affects snow crystal growth via the Gibbs-Thomson effect [2021Lib]. This gives the modified equilibrium vapor pressure of a spherical ice particle

$$c_{eq}(R) \approx c_{sat}(1 + d_{sv}\kappa) \quad (4)$$

where $d_{sv} = \gamma_{sv}/c_{ice}kT \approx 1\ nm$, $\gamma_{sv}$ is the solid/vapor surface energy, and $\kappa = 2/R$ is the curvature of the spherical surface. For a smooth but non-spherical surface, the curvature is defined as $\kappa = 1/R_1 + 1/R_2$, where $R_1$ and $R_2$ are the two principal radii of curvature of the surface. Note that this expression reduces to the normal flat-surface vapor pressure $c_{sat}$ when $R \to \infty$, as it must.

This small change in equilibrium vapor pressure can have a significant effect on snow crystal growth when the supersaturation is low and the surface curvature is high. Modeling reveals that the Gibbs-Thomson effect plays a important role in preventing the growth of unphysically thin plates at especially low supersaturations, as was demonstrated in [2013Lib1]. At high growth rates yielding dendritic structures, however, it appears that surface-energy effects are somewhat less important for determining morphological structures and growth rates.

## 6) Thermal diffusion

Latent heating produces an additional thermal-diffusion problem in snow crystal growth, as the resulting heat must be removed as the crystal grows. The double diffusion problem – particle and heat diffusion – can be solved analytically in the spherical case [2021Lib], revealing that particle diffusion is the dominant factor for growth in air. At low pressure, however, particle diffusion is fast and thermal diffusion becomes a limiting factor.

For snow crystal growth in air, thermal effects are generally negligible below -10 C, becoming progressively more important as one approaches the melting point [2016Lib]. Moreover, a rescaling of the far-away supersaturation $\sigma_\infty$ can be used to approximate the thermal effects to a reasonable approximation. For this reason, I will ignore latent heating and thermal diffusion for the remainder of this paper.

While the above list is not exhaustive, it does represent the primary physical processes that determine snow crystal growth rates and structure formation. Additional subtleties may reveal themselves as our understanding of this phenomenon improves, but for now we will restrict ourselves to the first five items presented above.

## ❄ A "STARTER" MODEL OF SNOW CRYSTAL GROWTH DYNAMICS

At present, it is too soon to consider a fully comprehensive numerical model of snow crystal growth. Our grasp of the underlying molecular dynamics of ice crystal growth is not complete, and no one has yet demonstrated numerical techniques that can reproduce realistic dendritic structures in the presence of strongly anisotropic attachment kinetics, at least not over a broad range of physical inputs. Instead, our initial objective here is to take the next small steps toward reaching this eventual goal.

We begin by suggesting a "starter" model of snow crystal growth dynamics that could be incorporated into a suitable numerical code. By comparing model calculations with laboratory observations, one imagines that further refinements of both the numerical strategies and the underlying growth physics may become apparent. Moreover, a general survey of growth behaviors and growth rates may suggest targeted experiments that greatly improve our current understanding of the various physical processes that guide snow crystal formation.

Focusing on growth in normal air, we define our starter model (call it M1) to have a normal growth velocity parameterized by

$$v_n = \alpha v_{kin} \sigma_{surf} \qquad (5)$$

with

$$\alpha(T, \sigma_{surf}) = e^{-\sigma_0/\sigma_{surf}} \qquad (6)$$

Our "best current guess" for the nucleation parameter $\sigma_0(T)$ is shown in Figure 1 for the basal and prism facets, these curves being defined by

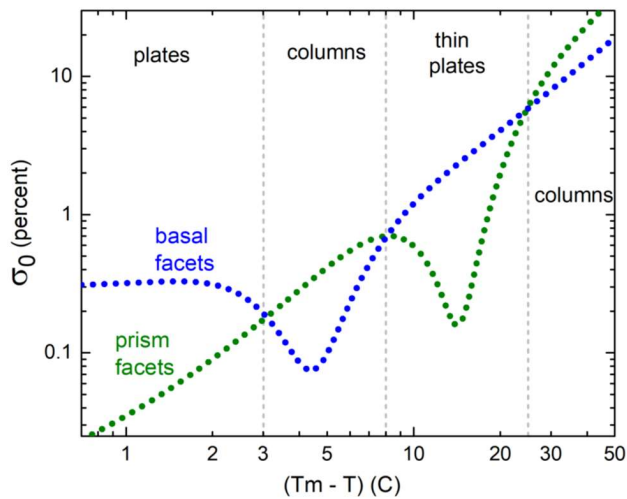

Figure 1: A plot of the nucleation parameter $\sigma_0(T)$ for both the basal and prism facets in the M1 model. The functional forms of both curves are defined in the text. Both are derived from crystal growth measurements including the SDAK phenomenon [2021Lib, 2019Lib1]

$$\sigma_{0,basal}(T) = (0.02T^{1.75} + 0.3) * \\ (1 - 0.87 \exp(-((\log(T) - \log(4.5))^2)/0.07)$$

$$\sigma_{0,prism}(T) = (0.015T^2 + 0.02T^{0.6}) * \\ (1 - 0.95 \exp(-((\log(T) - \log(14.4))^2)/0.06)$$

with temperature here in degrees Celsius. For all non-faceted surfaces, the M1 model assumes $\alpha = 1$.

While these functional forms are completely ad hoc, the values of $\sigma_0(T)$ derive mainly from experimental data. The overall shapes of the curves come from measurements of the growth of faceted ice crystals in near vacuum [2013Lib, 2021Lib], and we have additionally added the "SDAK dips" presented in [2020Lib1, 2020Lib2, 2021Lib].

To keep M1 relatively simple, we chose to set $A = 1$ in Equation 3 for all growth conditions, even though our data suggest that this is not entirely accurate for broad prism facets at high temperatures [2013Lib, 2021Lib]. In air, however, narrow prism facets do exhibit $\alpha \to 1$ at high growth rates [2020Lib]. Setting $A = 1$ for all prism facets should not yield horribly inaccurate results, therefore, and it substantially reduces the overall complexity of this starter model. The physical nature of the attachment kinetics on prism facets above -3 C remains somewhat puzzling, so it seemed prudent not to overthink M1 too much in this growth regime.

In addition to this model for the attachment kinetics, surface energy effects must also be included in the numerical calculations to incorporate the Gibbs-Thomson effect. This is necessary to prevent the growth of one-pixel-wide plates and other unphysical structures, as demonstrated in [2013Lib1]. It is not necessary, however, to calculate the local surface curvature to great accuracy in the model. Even a rough estimate of curvature is likely sufficient to suppress most unphysical structures. This reflects the fact that surface energy is not an especially important factor in snow crystal growth dynamics.

## Improving the SDAK effect

In the formulation above, our starter model does not fully incorporate the SDAK effect, because M1 does not distinguish between broad and narrow facets. We made this choice because the Edge-Sharpening Instability (ESI) [2021Lib, 2017Lib] is quite efficient in air, quickly turning broad facets into narrow facets during growth. In the observations presented below, we see that many fast-growing snow crystal morphologies do not contain any broad faceted surfaces. In these cases, therefore, it is a reasonable first approximation to assume that all faceted surfaces are described by the attachment kinetics on narrow facets.

As the next step after M1, an improved model (call it M2) could incorporate an attachment kinetics that depend on the widths of the respective faceted surfaces. Once again, the exact nature of this width dependence is probably not extremely important, because the ESI tends to provide fast transitions from broad to narrow facets. And the final widths of the latter are mainly limited by attachment kinetics on fast-growth surfaces [2002Lib].

In M2, the $\sigma_0(T)$ for broad facets could then be defined by:

$$\sigma_{0,basal}(T) = (0.02T^{1.75} + 0.3)$$

$$\sigma_{0,prism}(T) = (0.015T^2 + 0.02T^{0.6})$$

which is the same as M1 except without the SDAK dips. It is not clear at this time whether switching from M1 to M2 would yield many substantial differences in growth rates or morphological developments. Model testing is necessary to explore this question, along with detailed comparisons between model calculations and experimental data.

As mentioned above, it is not possible at this time to simply choose the appropriate attachment kinetics for our model, because we do not yet know what is appropriate; our understanding of the underlying physics is not good enough, Instead, progress will likely be made via a "bootstrap" process, beginning with a best-guess parameterization of the attachment kinetics and making additional refinements based on how well the model predictions match growth experiments.

## ❄ Comparison with Laboratory Observations

While it is possible to create a computational model of snow crystal growth from theoretical considerations alone, the phenomenon is so complex that detailed comparisons with experimental data are essential for making real progress. To facilitate such comparisons, Figure 2 shows a set of 206 snow crystal growth observations as a function of temperature and water vapor supersaturation in air, each showing the morphology and size of an ice crystal forming on the tip of a slender c-axis ice needle after a known growth time. The dual diffusion chamber used to create these photographs is described in detail in [2021Lib2]. We estimate that the temperature uncertainties in these data are typically ±0.2 C, and that the water-vapor supersaturations are between 0.8 and 1.2 times the stated values. The supersaturation is especially difficult to measure to this accuracy in situ, so we rely on diffusion-chamber modeling to determine $\sigma_\infty$ surrounding the growing crystals, as described in [2021Lib2].

For each of these 206 observations, the temperature and supersaturation were first set to their desired values, and the temperature was frequently measured using a small thermistor that could be moved into the observation region for this purpose. When the environmental conditions were stable, a set of several c-axis ice needles was moved into the observation region, supported by a vertical wire substrate as described in [2021Lib2]. The wire was rotated to select a suitable crystal for observation, which was subsequently photographed as it grew. Often focus stacking was used to provide a better overall image quality, as the high numerical aperture of the microscope objective yielded a rather small depth-of-field in a single photograph.

Each composite photo (after focus stacking) shows a representative example selected from several growing crystals in each multi-needle set, or from multiple sets. Some subjective preference was given to well-formed crystals exhibiting good symmetry and growing on isolated ice needles. Growth times and image scales are provided in each panel in Figure 2.

It takes some experience to understand the 3D morphological structures from the 2D image projections shown in Figure 2, and I will try to provide 3D sketches derived from my own experience (after years of observing thousands of crystals from various angles under different conditions) in subsequent papers in this series. A few such morphological sketches are shown in [2021Lib2].

## A roadmap for better understanding snow crystal growth dynamics

My main purpose in publishing this paper is to make available the observational data shown in Figure 2, as this series of photographs establishes a needed extension of the well-known Nakaya diagram. In addition to providing morphological information, these observations allow quantitative evaluations of snow-crystal model predictions over a broad range of conditions. Using slender ice needles as seed crystals provides a well-defined initial condition, while revealing the emerging structural features in far greater detail than previous observations.

At present, it is not clear that any existing 3D computational models of snow crystal growth can adequately reproduce any of the structures seen in Figure 2 under prescribed growth conditions, let alone reproducing the entire set with even modest fidelity. However, the images immediately suggest a roadmap for developing better models: 1) develop a physical model of the ice growth dynamics, such as the M1 and M2 models suggested in the previous section; 2) develop an adequate computational system than can combine diffusion-limited growth with strongly anisotropic attachment kinetics; and 3) compare model predictions with the combined observations in Figure 2.

As this line of research progresses, observations like those in Figure 2 will likely lead to a better grasp of the physical processes guiding ice crystal growth, in addition to simply testing model predictions. Moreover, comparing models with observations should suggest additional targeted experimental investigations that can inform microphysical theories beyond their current state.

As our understanding of the molecular dynamics of crystal growth improves, and the surface dynamics of ice specifically, it should eventually be possible to connect a parameterization of the attachment kinetics with molecular dynamics simulations of the ice surface. Investigations of the facet step energies illustrate progress in this direction [2020Llo], and more sophisticated analyses will surely follow. In the (somewhat distant) future, it may be possible to establish an unbroken "chain of reasoning" that connects nanoscale molecular dynamics processes at the ice surface to parameterized models of the attachment kinetics, and finally to macroscopic snow crystal growth observations like those in Figure 2. This would represent an essentially full comprehension of the physical processes underlying snow crystal formation, providing an important example of how a complex and enigmatic physical phenomenon can be deciphered via reductionist scientific reasoning.

## ❄ References

Figure 2: (Following four pages) Laboratory observations of snow crystals growing on the ends of slender c-axis ice needles [2021Lib2]. Growth occurred in air under constant environmental conditions, and each of the 206 panels is labeled with the temperature, water-vapor supersaturation, growth time, and the physical size of the field-of-view of each square image.

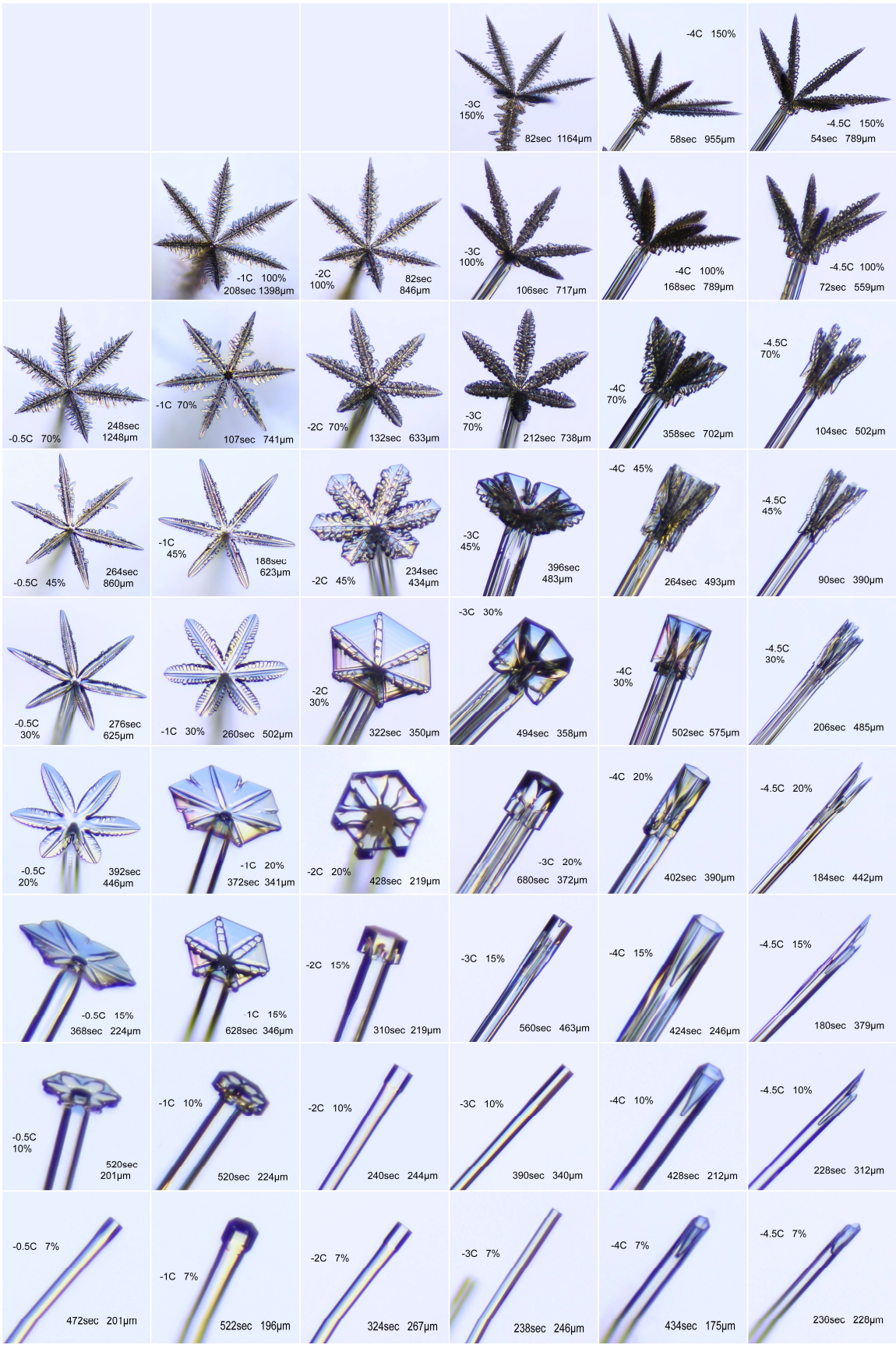
11

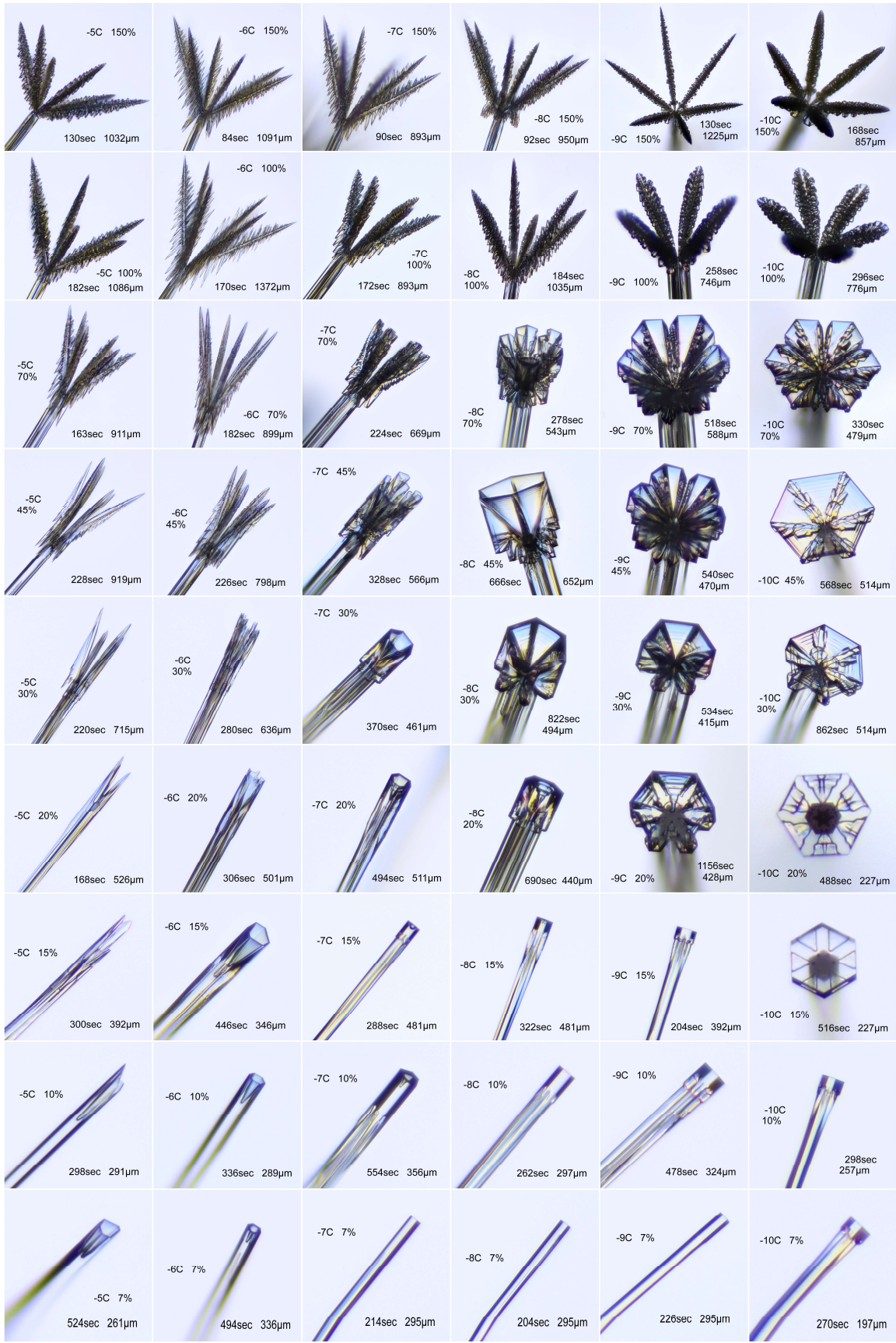

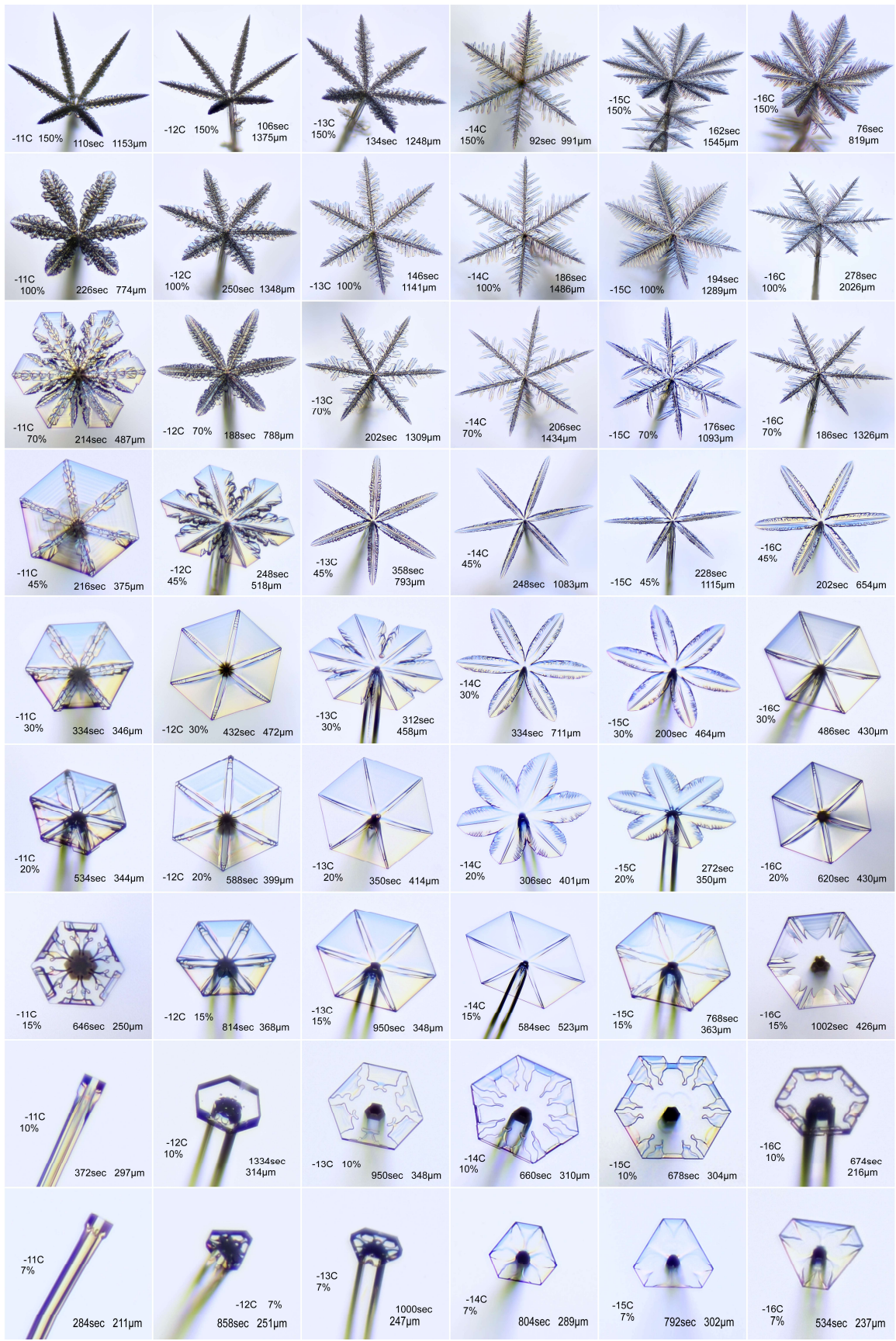

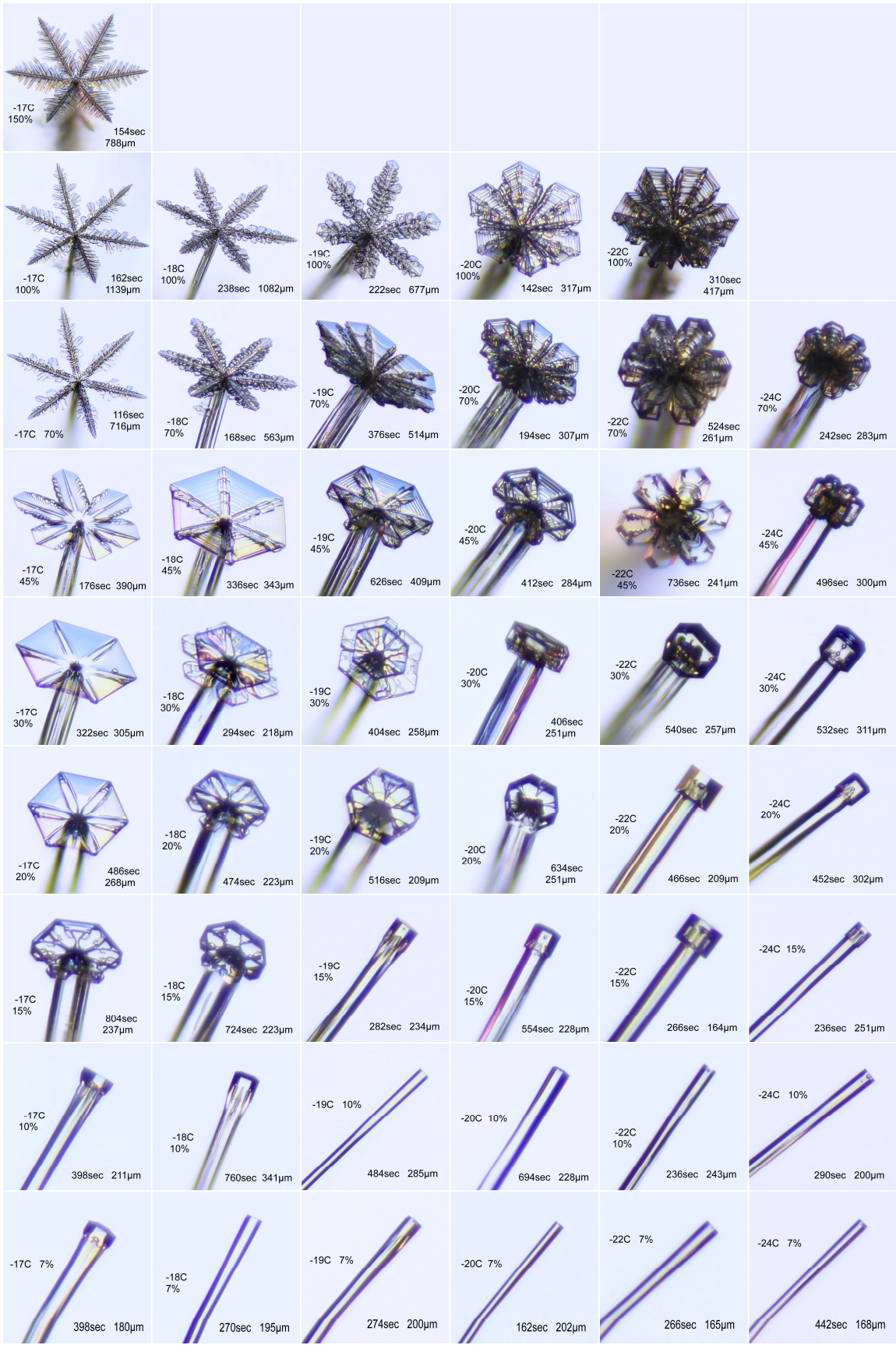
14